\documentclass{elsart}

 \usepackage{graphicx}
\usepackage{amssymb}

\begin{document}

\begin{frontmatter}

% Title, authors and addresses

% use the thanksref command within \title, \author or \address for footnotes:
% \title{Title\thanksref{label1}}
% \thanks[label1]{}
% \author{Name\thanksref{label2}}
% \thanks[label2]{}
% \address{Address\thanksref{label3}}
% \thanks[label3]{Via A.Valerio 2, I-341127 Trieste, Italy}
% including your email address:
% \address{Address\thanksref{email}}
% \thanks[email]{E-mail: paolo.milazzo@trieste.infn.it}

\title{
Size and asymmetry of the reaction entrance channel: 
influence on the probability of neck production
}

% use optional labels to link authors explicitly to addresses:
 \author[ts]{P.M.Milazzo},
 \author[bo]{G.Vannini},
 \author[lns]{C.Agodi},
 \author[lns]{R.Alba},
 \author[lns]{G.Bellia},
% \author[bo]{M.Bruno},
 \author[ba]{N.Colonna},
 \author[lns]{R.Coniglione},
% \author[bo]{M.D'Agostino},
 \author[lns]{A.Del Zoppo},
 \author[lns]{P.Finocchiaro},
 \author[lnl]{F.Gramegna},
 \author[mi]{I.Iori},
 \author[lns]{C.Maiolino},
 \author[ts]{G.V.Margagliotti},
 \author[lnl]{P.F.Mastinu},
 \author[lns]{E.Migneco},
 \author[mi]{A.Moroni},
 \author[lns]{P.Piattelli},
 \author[ts]{R.Rui},
 \author[lns]{D.Santonocito},
 \author[lns]{P.Sapienza}.

 \address[ts]{Dipartimento di Fisica and INFN, Trieste, Italy}
 \address[bo]{Dipartimento di Fisica and INFN, Bologna, Italy}
 \address[lns]{INFN, Laboratori Nazionali del Sud, Catania, Italy}
 \address[ba]{INFN, Bari, Italy}
 \address[lnl]{INFN, Laboratori Nazionali di Legnaro, Italy}
 \address[mi]{Dipartimento di Fisica and INFN, Milano, Italy}

\begin{abstract}
The results of experiments performed to investigate the Ni+Al, Ni+Ni, Ni+Ag
reactions at 30 MeV/nucleon are presented. From the study of
dissipative midperipheral collisions, it has been possible to detect events
in which Intermediate Mass Fragments (IMF) production takes place.
The decay of a quasi-projectile has been identified; its
excitation energy leads to a multifragmentation totally described in terms of
a statistical disassembly of a thermalized system
(T$\simeq$4 MeV, E$^*\simeq$4 MeV/nucleon). Moreover, for the systems 
Ni+Ni, Ni+Ag, in the same
nuclear reaction, a source with velocity intermediate between that of the
quasi-projectile and that of the quasi-target, emitting IMF, is observed. 
The fragments produced by this source are more neutron
rich than the average matter of the overall system, and have a 
charge distribution different, with respect to those statistically emitted 
from the
quasi-projectile. The above features can be considered as a signature of the
dynamical origin of the midvelocity emission. The results of this analysis
show that IMF can be produced via different mechanisms simultaneously present
within the same collision. Moreover, once fixed the characteristics of the
quasi-projectile in the three considered reactions (in size, excitation 
energy and temperature), one observes that the probability of a partner IMF 
production via dynamical mechanism has a threshold 
(not present in the Ni+Al case) and increases with the size of the 
target nucleus.
\end{abstract}

\begin{keyword}
% keywords here, in the form: keyword \sep keyword
Heavy Ions \sep Multifragmentation
% PACS codes here, in the form: \PACS code \sep code
\PACS 25.70.Pq \sep 25.70.-z
\end{keyword}
\end{frontmatter}

% main text
%\section{}
%\label{}
%\begin{multicols}{2}
%\twocolumn

\section{INTRODUCTION} 
The production of intermediate mass fragments (IMF, Z$\ge$3) is one 
of the main features of the nuclear reactions in the Fermi energy regime 
(i.e. at bombarding energies of 30-50 MeV/nucleon), and can arise from
various mechanisms \cite{libri}. 

Compound systems, formed in central collisions, break into several IMFs.
This behaviour has been described in terms of a statistical approach in which 
low density nuclear matter is supposed to have a liquid-gas phase transition 
\cite{lg}. In fact the experimental observables, charge distribution and 
partition, 
and the shape of the caloric curve (temperature versus excitation energy) 
\cite{exp-lg} are in good agreement with the predictions of such statistical 
multifragmentation models \cite{smm}. 

At these energies in peripheral and midperipheral collisions, it has been 
observed that the quasi-projectile (QP) and the quasi-target (QT), 
can de-excite following a statistical pattern and giving rise to the 
production of IMF. 

On the other hand, many experiments have shown that at mid-rapidity 
dynamical mechanisms lead to the production of IMF; this effect is due to the 
rupture of a neck-like structure joining QP and QT \cite{exp-neck,neck1}. 
Various transport calculations predict that dynamical fluctuations dominate 
the neck instability allowing the production of IMF \cite{theo-neck}; 
moreover the experimental results (in particular concerning the 
charge distribution and the isotopic composition of fragments) can not be 
described in terms of statistical approaches. 

It has been shown that in midperipheral collisions it is possible to observe 
inside the same 
event the competition between statistical and dynamical mechanisms 
leading to the production of IMF \cite{neck1}. 

To better investigate this phenomenon we experimentally studied the 
Ni+Al, Ni+Ni, Ni+Ag midperipheral collisions at 30 MeV/nucleon. The 
results of this investigation are presented and discussed in this paper. 

At first, within the same set of mid-peripheral events, we separate
the IMFs coming from the statistical 
disassembly of the QP from those coming from a dynamically driven neck 
rupture. Then, we study the balance between these two 
mechanism of IMF production for the three different interacting systems. 
The comparison between the IMF produced via statistical and dynamical 
processes show significant differences concerning the charge distributions 
and the isotopic composition of the fragments. 
The analysis of the different systems will demonstrate that the neck
formation probability is strongly influenced by the size of the target. 

In Sect.2 a description of the experimental conditions is given; 
the mid-peripheral collisions features are discussed in Sect.3; 
Sect.4 is devoted to the analysis of the QP emitting source formed 
in the three different reactions studied. The production of IMF at 
midvelocity is discussed in Sect.5, then the conclusions are 
drawn in Sect.6. 

\section{EXPERIMENTAL SET-UP AND DATA ANALYSIS PRESCRIPTIONS }
The experiment was performed at the INFN Laboratori Nazionali del Sud, where 
the superconducting cyclotron delivered a beam of $^{58}$Ni at 
30 MeV/nucleon, using the MEDEA \cite{Medea} and MULTICS \cite{Multics} 
experimental apparata as detectors.
The angular range 3$^{\circ}<\theta_{lab}<$28$^{\circ}$ 
was covered by the MULTICS array \cite{Multics}, which
consists of 55 telescopes, each made of an Ionization Chamber (IC), a Silicon 
position-sensitive detector (Si) and a CsI crystal. 
The typical values of the energy resolutions
are 2\%, 1\% and 5\% for IC, Si and CsI, respectively.
The identification threshold in the MULTICS array was about 1.5
MeV/nucleon for charge identification. 
Good mass resolution for light isotopes (up to Carbon) was obtained. 
Energy thresholds for mass identification of 8.5, 10.5, 14 MeV/nucleon
were achieved for $^4$He, $^6$Li and $^{12}$C nuclei respectively.
The 4$\pi$ detector MEDEA is made of 180 Barium 
Fluoride detectors placed at 22 cm from the target and it
can identify light charged particles (Z=1,2) (E$\leq$300 MeV) 
and $\gamma$-rays up to E$_\gamma$=200 MeV in the 
polar angles from 30$^{\circ}$ to 170$^{\circ}$ and in the 
whole azimuthal angle \cite{Medea}. 

In these experiments light charged particles and fragments were 
detected on an event by event basis, thus allowing the  
description of the reaction dynamics.

In heavy ions reactions at intermediate energies different decaying systems 
are formed, depending on the impact parameter, and become the source
of fragments which differ in size, shape, excitation energy, 
and in the way they are formed. Therefore one must identify the decaying 
systems and ensure that all the fragments are correctly
assigned to one of these systems. Thus, since the aim of this paper is 
to present data on IMF production in the following we will restrict our 
analysis only on many-fragments events \cite{neck1}. 
Since however many fragments can be produced both in central and midperipheral 
collisions, it is mandatory to distinguish collisions 
occurred at different impact parameters, in order to have a comprehension of 
the mechanisms responsible for IMF production and emission 

The impact parameter data selection is based on the heaviest 
fragment velocity. We can select peripheral and 
midperipheral events when the heaviest fragment (produced by the disassembly 
of a QP emitting source) in the laboratory frame travels at velocities 
higher than 80\% of that of the projectile (v$_P$=7.6 cm/ns); 
on the contrary, in central collisions the heaviest fragment
travels at velocities close to that of the centre of mass.
Only ``complete'' events are analyzed, i.e. when at least 3 IMF are
produced (with the heaviest fragment having Z$\geq$9) and more than 
80\% of the total linear momentum is detected. Accordingly, since the energy 
thresholds prevent from detecting the QT reaction products, we 
find that the total detected charge (Z$_{Tot}$) does not differ from that of 
the projectile for more than 30\% (20$\leq$Z$_{Tot}\leq$36). 

\section{DYNAMICAL AND STATISTICAL IMF PRODUCTION IN MID-PERIPHERAL 
COLLISIONS }
The results presented hereafter will refer only to mid-peripheral 
collision events, with at least three detected IMF,
observing the IMFs emitted from the QP and from the mid-velocity
neck (neck-IMF in the following), and studying the different and competitive 
IMF production mechanisms.
 
In Fig.1 the yields of carbon and oxygen fragments (for the three 
considered reactions) are plotted as a function of the 
component of the velocity parallel to the beam axis. 
The centre of mass velocities for the three systems are 5.18 (Al), 3.80 
(Ni) and 2.65 (Ag) cm/ns.

The IMF possibly coming from the QT and part of those having mid-velocity 
were not detected. The problem affects the study
of mid-velocity IMF mainly for the Ni+Ag reaction.

Beginning with the upper panels of Fig.1 (Ni+Al), at the centre of mass
velocity there is a minimum in the production of Z=6-8 fragments;
this fact suggests a negligible formation of a neck-like structure for
this light system.
On the contrary, in Ni+Ni we notice that at 
mid-velocity a large contribution of IMF is present \cite{neck1}.
At last, the lower panels (Ni+Ag) show evidence of a larger
contribution of mid-velocity IMF (even if there is a
clear efficiency cut).

\begin{figure}[htbp]
\begin{center}
\includegraphics[width=10cm,height=10cm]{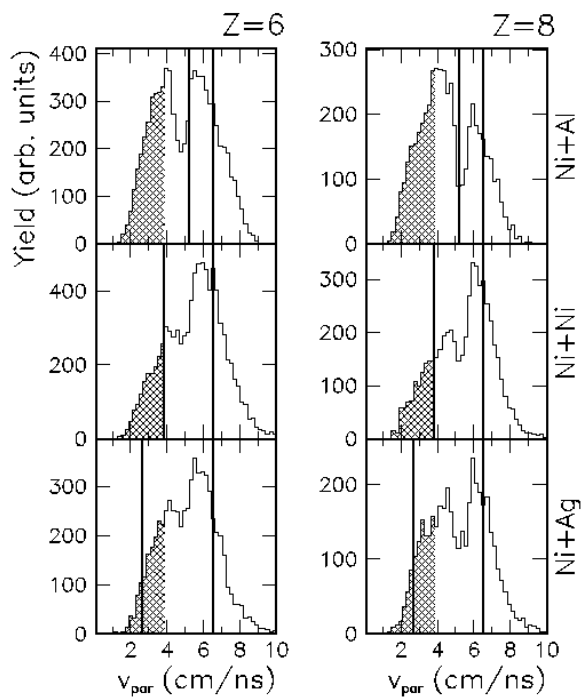}
\end{center}
\caption{
Experimental v$_{par}$ distributions for Z=6 (left panels) and Z=8 
(right panels), for the three studied reactions; vertical lines refer to 
the center of mass and QP velocities. Experimental efficiency cut occurs 
in the shadowed area.}
\end{figure}

Thus, for mid-peripheral collisions, while the disassembly of a 
QP (and a QT) is present in all the three considered systems,
the production of IMF at mid-velocity depends on the size 
of the target nucleus. 

\section{THE IMF EMITTED FROM THE QP DECAY }
To compare the reaction mechanisms for midperipheral collisions for the 
three different interacting systems we have to select a set of 
``complete'' events (as described in Sec.2) for which the QPs have very 
similar characteristics; then, we will study the process leading to its
disassembly. We will further restrict the analysis to fragments emitted 
with v$_{par}>$ 6.5 cm/ns (QP-IMFs in the following), forcing the
selection of the QP decay products forward emitted (see for instance Fig.1),
with negligible contamination due to QT and midvelocity source emission. 

In order to evaluate the degree of equilibration reached by the QP 
before its disassembly, we measured the angular and energy distribution of 
the QP emitted isotopes, in their reference frame. 

\subsection{THE QP-IMFs ANGULAR AND ENERGY DISTRIBUTIONS}
The investigation of the angular distributions is also aimed at verifying
if the QP fragments are produced by a nearly isotropic emitting source as 
expected for a statistical decay. 
The angular distributions of QP fragments for the three 
reactions are presented in Fig.2; the flat shape is
in agreement with the hypothesis of an isotropic emission, a
necessary condition to establish a possible equilibration of the studied
system.

\begin{figure}[hbtp]
\begin{center}
\includegraphics[width=10cm,height=10cm]{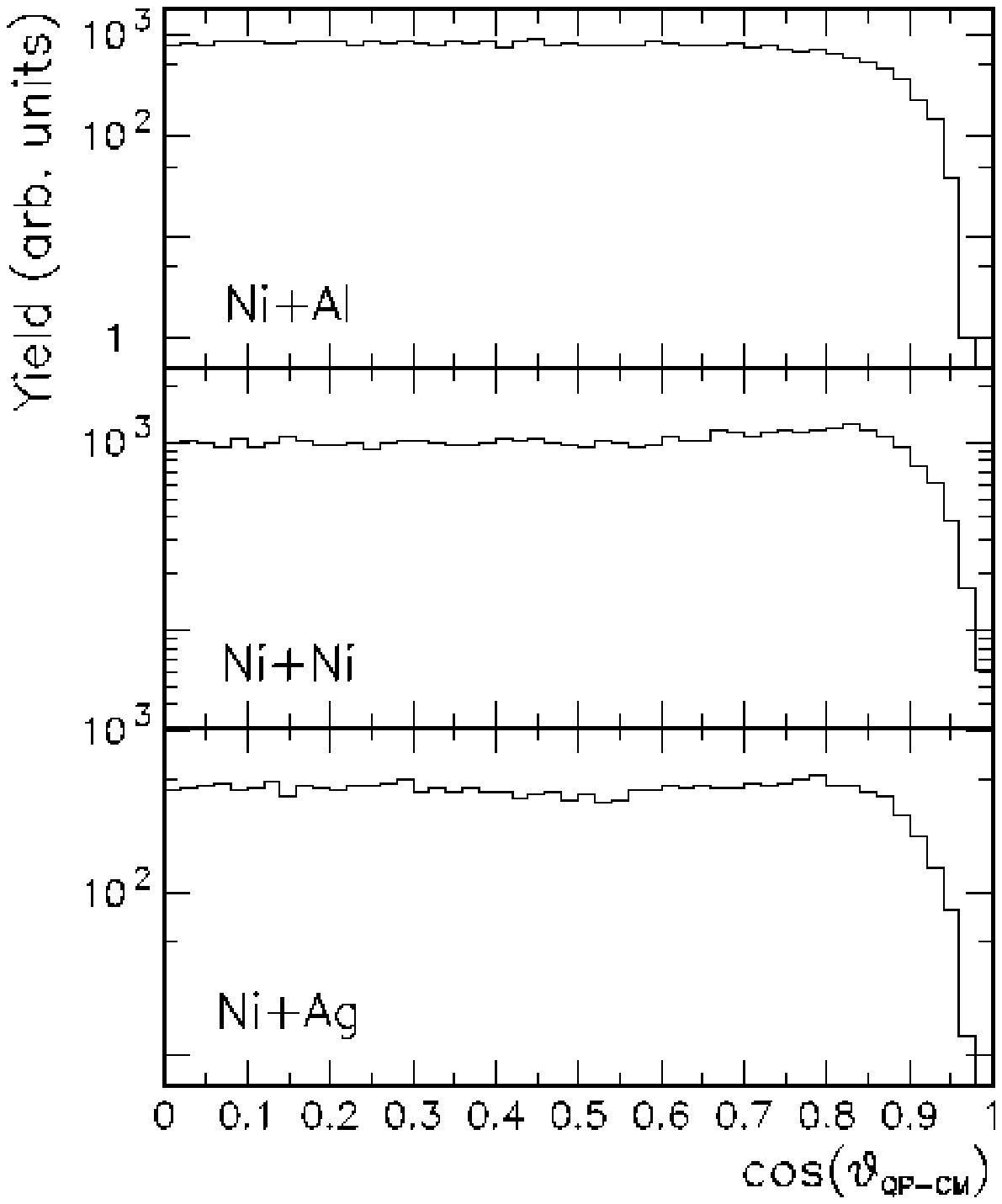}
\end{center}
\caption{
Angular distributions for IMF forward emitted by the QP, for 
the three studied reactions}
\end{figure}

Energy distributions can be strongly influenced by the fact that 
Coulomb and collective energies are mass dependent;
energy spectra of different isotopes may display different slopes \cite{Bauer}.
On the contrary, the thermal energy contribution must
be the same for all masses; by fitting the energy distributions
with a Maxwellian function (for a surface emission)
\begin{equation}
Y(E)={(E-E_0)\over T_{slope}^2}\cdot e^{-(E-E_0)\over T_{slope}} 
\end{equation}
we find comparable values of $T_{slope}$ for all the detected isotopes
(3$\leq$A$\leq$14). $T_{slope}$ is the parameter related to the
apparent temperature, and $E_0$ is a parameter related
to the Coulomb repulsion. The results are reported in Table I. 

The behaviour of angular and energy distributions indicates that 
the condition of equilibration of the fragmenting QP systems is satisfied. 

From the comparison of the QP behaviour in the 
three different reactions, it is possible to notice the 
similarity of the obtained apparent temperature 
slopes, independent from the considered isotope. 
As an example in Fig.3 the energy distribution of $^6$Li and $^{10}$B isotopes 
are compared; the results of the maxwellian fit are superimposed.

\begin{figure}[htbp]
\begin{center}
\includegraphics[width=10cm,height=10cm]{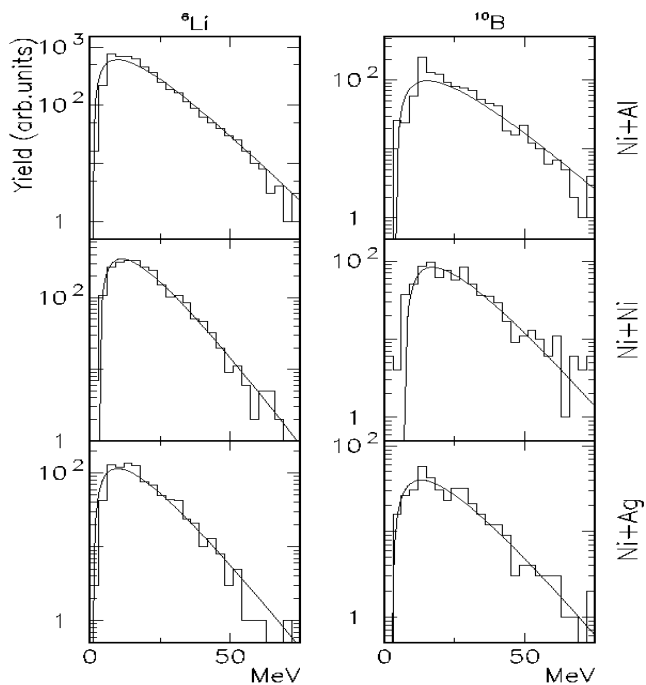}
\end{center}
\caption{
Energy distributions for the $^6$Li and $^{10}$B isotopes; 
maxwellian fits are superimposed}
\end{figure}

\subsection{THE QP-IMFs CHARGE DISTRIBUTION}
The following point is related to the study of the QP-IMFs charge 
distributions, presented in Fig.4. 
We have to stress that, with the adopted data selection, 
the distributions are quite similar; the QP mean elemental 
charge multiplicities of the fragments, produced 
in the Ni+Al and Ni+Ni cases, are overlapping, and
the difference presented by the Ni+Ag case at large values of Z,
is probably due to a smaller excitation 
energy of this QP or to a pick-up of few nucleons from the target. 

\begin{figure}[htbp]
\begin{center}
\includegraphics[width=10cm,height=10cm]{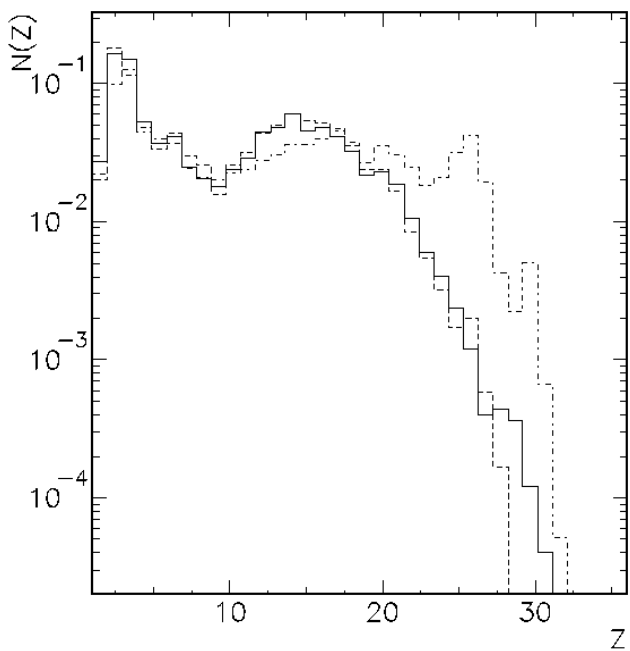}
\end{center}
\caption{
Mean elemental event multiplicity N(Z) for QP charged products 
(full line Ni+Al, dashed line Ni+Ni, dot-dashed line Ni+Ag)}
\end{figure}

\subsection{THE QP-IMFs ISOTOPIC COMPOSITION}
Isotopic effects in nuclear reactions have recently received attention because 
of their relation with the simmetry energy in the nuclear equation of state 
\cite{isospin}.

Even thought the QP-IMF charge distributions present a similar shape in the 
three considered cases, the isotopic composition of fragments could be 
affected by the different N/Z ratio of the three different targets. 
In Table II are reported the measured isotopic composition of the QP-IMFs, 
expressed in percentage terms of the yields ratio Y(Z, A)/Y(Z), for fixed 
Z values. No significant fluctuations can be appreciated among the three 
analysed reactions. 
It is important to stress that many experimental evidences have shown that the 
neck IMFs, reaction partner of the studied QPs, are neutron rich 
\cite{exp-neck,neck1,neck2}. 
However, the QP characteristics result unchanged, with respect to those of the 
starting Ni projectile nucleus. To this point in Table III are presented the 
average values of the N/Z ratios at different Z, and they are close
to the value (1.07) of the projectile Ni nucleus, and very similar to those 
of the stable nuclei. 

\subsection{THE QP EXCITATION ENERGY AND TEMPERATURE}
Since energy and angular distributions satisfy some necessary conditions that
support the hypothesis that the QP has been subject to an
equilibration process, we can investigate some of its thermodinamic
characteristics (temperature and excitation energy).

For this experiment it is not possible to perform an evaluation
of the excitation energy through calorimetry \cite{calo},
because this technique
requires a careful event by event assignment of each fragment to its emitting 
source\cite{vient}, and here it is not possible due to the
overlap of distributions between midvelocity and QP velocity.
The excitation energies were therefore estimated by comparing the data with the
SMM predictions \cite{smm} which best describe the experimental findings of 
the QP fragment emission. 
In Ref.\cite{mangiarotti} it is shown that quanto-molecular dynamics 
calculations suggest that the QP size doesn't differ significantly from 
that of the projectile. The calculations were then performed for a Ni
nucleus at one third of the normal density. 
The events generated by SMM for different input excitation energies
were filtered with the experimental constraints. 
Experimental charge distributions were better reproduced
by choosing an excitation energy of 4.0, 4.0 and 3.5 MeV/nucleon for the
decaying QP in the Ni+Al, Ni+Ni and Ni+Ag, respectively
(see for instance Fig.8 of Ref.\cite{neck2}).

The temperature was evaluated by means of the double ratios of isotope 
yields \cite{albergo}.
The double ratio $R$ of the yields $Y$ of four isotopes in their ground states,
prior to secondary decay is given by:
\begin{equation}
R= {Y(A_1,Z_1)/Y(A_1+1,Z_1)\over Y(A_2,Z_2)/Y(A_2+1,Z_2)}=
{e^{B/T}\over a} \label{1}
\end{equation}
where $a$ is a constant related to spin and mass values and

$B=BE(Z_1,A_1)-BE(Z_1,A_1+1)-BE(Z_2,A_2)+BE(Z_2,A_2+1)$,

and $BE(Z,A)$ is the binding energy of a nucleus with charge $Z$
and mass $A$.

In principle, $R$ gives directly the
temperature $T$. However, primary fragments can be excited
so that secondary decays from higher lying states of the same
and heavier nuclei can lead to non-negligible distortions of
the measured ratios $R$. 
In Refs.\cite{Betty,Betty-MSU} an empirical procedure was proposed,
to strongly reduce such distortions; 
it was shown \cite{Betty-MSU,xecu} that for temperatures near 4 MeV these 
empirical correction factors do not depend
either on the size or on the N/Z ratio of the decaying systems.

Moreover to apply the double ratios method \cite{albergo}
one has to be sure that the nuclei originate from
the same emitting source and therefore, when the contributions of different
sources are present, particular care must be taken in selecting the isotopes.

The break-up temperatures $T$ of the QP decaying system were extracted 
averaging the values obtained from different double ratios of isotope 
yields, corrected as suggested in Ref.\cite{Betty}. 
The experimental temperatures and excitation energies of the present 
measurements, are reported in Table IV.

\subsection{THE QP CHARACTERISTICS SUMMARY}
In summary the adopted data selection allowed us  
to select the mid-peripheral collisions for which, in the exit channel 
of the reaction, the QP decaying systems have the same characteristics, in 
the three considered reactions. 

In particular, the QP disassembly is well described within a statistical 
framework, and its properties are: a) a size close to that of 
the incident Ni nucleus, b) an excitation energy around 4 MeV/nucleon, c) a 
temperature around 4 MeV. Properties b) and c) place the system well inside 
the plateau of the caloric curve, where statistical multifragmentation is 
the main decay pattern. 

\section{THE MIDVELOCITY IMF PRODUCTION}
From what shown in the previous chapter it is clear that we are working with a 
particular channel in which (changing the target) we have the same 
excitation energy for the QP nuclei. 
Therefore, it is interesting to investigate what happens to the midvelocity 
IMF production, partner in the reaction, in the three different cases. 
In fact, once fixed the energy dissipation process, the IMF neck 
production can be directly related to the different size and asymmetry of 
the entrance channel. 

As shown in Fig.1 a large amount of IMFs are emitted at midvelocity in 
midperipheral Ni+Ni and Ni+Ag reactions; contribution from such emission is 
not sizeably present in the lighter analysed system Ni+Al. While for all the 
three 
reactions the fragmentation of the QP is very well explained in terms of 
statistical break-up, the presence, inside the same event, of IMF at 
intermediate velocity, can not be explained in terms of a pure statistical 
theory; different studies \cite{exp-neck} have shown that the origin of 
midvelocity 
IMF can be considered of dynamical nature. In particular, by comparing 
the characteristics of the statistically emitted QP-IMFs and those of the 
neck IMFs, it is possible to observe how their charge distribution and 
isotopic composition are significantly different \cite{neck1,neck2}; 
these evidences fortify the idea that two competitive reaction 
processes can take place simultaneously. 

The characteristics of neck IMFs have been evaluated by means of fit
procedures, that rely on the fact that the QP properties are well established;
the characteristics of the neck IMFs are then extracted studying the 
deviations from statistical 
distributions, as described in the following. 
The main assumption is that fragments emitted with velocities higher than that 
of the QP (v$_{QP}>$6.5 cm/ns) origin only from the QP decay (forward 
emission from the QP, with negligible contribution coming from other 
source disassemblies). 

We fitted the QP-IMFs velocity distributions taking into account only the
forward emission region, by means of a gaussian function with 
its maximum fixed at the QP velocity. This procedure was repeated for each 
fragment charge in the range Z=3-14 (see for instance at Fig.9 of 
Ref.\cite{neck2}). From the results it was then possible to extract the 
yield Y$_{QP}$(Z) for each fragment emitted by the QP. 

Due to the experimental energy threshold the velocity spectra are affected by 
detection inefficiencies. Then, we restricted our analysis to velocities 
higher than 3.8 cm/ns, where the distributions are not influenced by 
experimental cuts. This value has been chosen because the IMF 
emitted from the QT decay can not have velocities (in the laboratory frame)
that exceed 3.5 cm/ns (this was checked by using the predictions of the 
Classical Molecular Dynamics model \cite{cmd}). 
The yield of the neck IMFs (Y$_{Neck}$ contribution) has been extracted by 
means of a two emitting sources fitting procedure: one source is related to 
the QP, and its parameters were completely determined in Sec.4,
the other is centered at the centre of mass 
velocity and takes into account the midvelocity fragments. 
In the fitting procedure we used two gaussian distributions to reproduce the 
experimental data; there is not a physical reason to justify this 
choice for the dynamical 
component, however the results are not affected by this particular constraint. 
In fact, no differences were found between the present results and those 
already published \cite{neck1,neck2} obtained for the Ni+Ni mid-peripheral 
collisions, where a direct quantitative analysis was possible. 

The results for the Ni+Ag reaction are presented in Fig.5. 

\begin{figure}[htbp]
\begin{center}
\includegraphics[width=10cm,height=10cm]{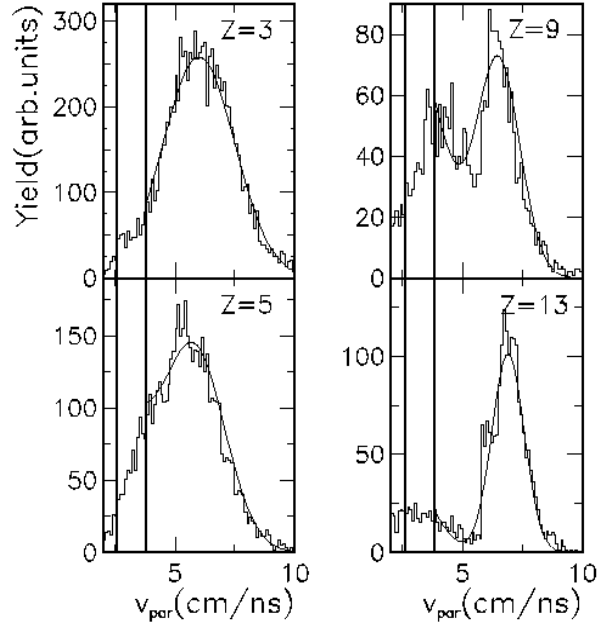}
\end{center}
\caption{
Experimental Ni+Ag $v_{par}$ distributions (Z=3, 5, 9, 13)
and superimposed fit; the lines refer to 
the centre of mass (2.65 cm/ns) and efficiency threshold (3.8 cm/ns)}
\end{figure}

The comparison between the total and QP-IMFs $v_{par}$ distributions allows 
us to evaluate the yield (Y$_{Neck}$) at midvelocity. 

The IMF charge distributions (from QP and midvelocity) are 
very different: the IMF coming from a neck rupture 
mainly have charges between that of carbon and of oxygen. 
In order to enhance this aspect, in Fig.6a the ratio between the relative 
yields (Y$_{Neck}$/Y$_{QP}$) is presented (for the Ni+Ni and Ni+Ag cases) 
as a function of the atomic number Z. We observe a bell-like shape, with very 
similar behaviour in both reactions. 
The fact that the maximum of this ratio is located at Z=9 is due to the 
strong decrease of the QP charge distribution in this region (see Fig.4).
It is worthwhile to notice the higher amount of neck IMFs produced in the 
Ni+Ag reaction.
In Fig.6a the relative yield (Y$_{Neck}$/Y$_{QP}$) 
for the Ni+Ni reaction is multiplied by a factor 1.862 (which is the 
ratio between the Ag and Ni mass (108/58)); we 
observe that, except for the two lighter and less probable neck IMFs, the 
double ratio between the relative yields is almost constant (Fig.6b) 
around the value 108/58=1.862. 
This fact suggests that the size of the target nucleus plays a direct role in 
the amount of neck IMFs production. 

\begin{figure}[htbp]
\begin{center}
\includegraphics[width=10cm,height=10cm]{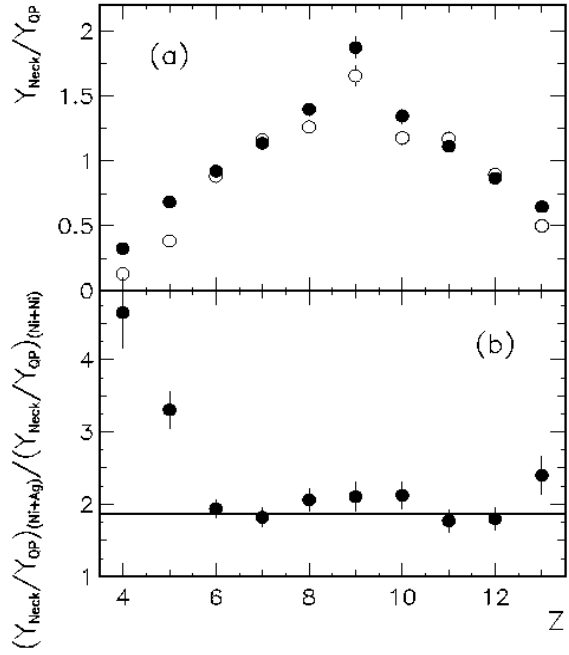}
\end{center}
\caption{(a) Upper panel: ratio of the measured yield for neck fragmentation 
and QP emission (open points Ni+Ni, multiplied by a factor 1.862; 
full points Ni+Ag); (b) lower panel: double ratio between relative yields}
\end{figure}

In many references \cite{exp-neck,neck1,neck2} it has been shown that the IMFs
coming from a neck like structure differ from those produced in a QP decay 
for what concern the isotopic composition. In Fig.7 the relative yields of 
different isotopes are presented; it is clear that the neck IMFs are 
heavier in mass (for fixed Z values) than those emitted by the QP. 

\begin{figure}[htbp]
\begin{center}
\includegraphics[width=10cm,height=10cm]{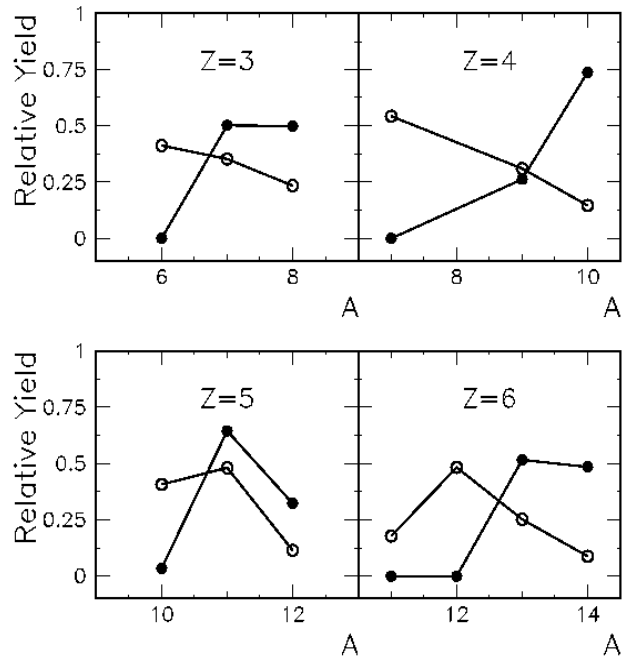}
\end{center}
\caption{Relative yields of different isotopes for fragments with 
charges from Z=3 to Z=6, for the Ni+Ni reaction. 
Open circles are related to the QP-IMFs, full circles represent the neck IMFs 
}
\end{figure}

The energy threshold to extract the mass value of the detected fragments is
higher than that allowing for charge identification; this experimental 
inefficiency does not permit a quantitative investigation of the 
isotopic composition of neck IMFs in the Ni+Ag reaction 
(no information is available on the mass 
of the fragments with velocity lower than $\approx$5 cm/ns). 
However, since the QP-IMFs forward emitted distributions are not affected by 
experimental cuts, we can give a qualitative evaluation of the isotopic 
composition of neck IMFs, looking at the very poor available data in the 
QP backward side (searching information by means of the comparison QP 
forward-backward emission). 
In Table V the average values of the N/Z ratio for different Z numbers are 
presented. The energy threshold increases with the mass of the 
detected nucleus. 
The higher identification thresholds for the heavier isotopes weakly affect 
the calculation of the average values of the N/Z ratio, for ther Ni+Ag case. 
Then for this system, to be conservative, we can give only a lower limit for 
this ratio. 

Then, not only the 
charge distribution, but also the isotopic composition of the neck IMFs
is very similar in the two analised systems. 

In summary the IMFs emitted in dynamical processes such as the neck formation 
have a charge distribution and neutron contents which are very different from 
those emitted in statistical processes. In particular the isospin composition 
of the dynamically emitted IMFs is very different even from that of the total 
system. 
\section{CONCLUSIONS}
The experimental investigation of the reactions Ni+Al, Ni, Ag 30 MeV/nucleon 
were performed at 
the Superconducting Cyclotron of the INFN Laboratori Nazionali del Sud, 
Catania.

In the study of the dissipative midperipheral  
collisions it has been possible to investigate the characteristics of IMF 
produced by two different types of reaction mechanisms.
The data analysis prescriptions for the impact parameter selection
allowed to select a well defined set of events;
in the study of the Ni+Ni and Ni+Ag reactions it has been possible to select 
events in which the IMFs are competitively emitted by the decay of the QP 
and by an intermediate velocity source.

Concerning the disassembly of the QP it has been verified that this system
reaches a thermal equilibrium before decaying following a statistical pattern.
This point was clarified looking at the experimental angular and
energy distributions; isotropic angular distributions and maxwellian shape
for the energy distributions give an indication that the thermalization
has taken place. A comparison with the SMM predictions strongly support
this hypothesis. 

The analised QP systems present the same characteristics in the three 
considered reactions; in particular their temperature and excitation energy 
(T$\simeq$4 MeV, E$^*\simeq$4 MeV/nucleon) suggest the multifragmentation as 
the main statistical de-excitation channel. 

Inside the same nuclear events, in the Ni+Ni, Ni+Ag collisions, IMF production 
is present also at midvelocity, due to dynamical processes.
On the contrary, the Al target seems to be too light to allow the formation 
of a neck structure, from the overlap of projectile and target during the
collision. 

The neck IMFs, when compared to the products of the QP decay, show a 
very different behaviour for what concern the 
charge distribution and the isotopic content of the fragments.
These evidences are taken as a signature of the different nature of the 
two processes, statistical and dynamical, leading to the formation of IMF. 

The charge distribution and the average values of the N/Z ratio (for different 
Z numbers) of the neck IMFs produced in the Ni+Ni and Ni+Ag reactions are 
very similar. 

Once fixed the QP characteristics (size, excitation energy and 
temperature) and verified that the partner dynamical IMF production present 
similar features in different reactions, we observe that the production amount 
of neck IMFs increases with the size of the target nucleus. 
\newpage
\begin{table}
%\begin{center}
%{\bf Table I}
%\end{center}
\caption{
Temperature parameters extracted from a Maxwellian 
fit procedure of the isotope energy spectra (typical fit error on 
extracted values is $\pm$1 MeV)}
  \begin{tabular}{lllll}
\hline
$Z$&$A$&$T_{slope}$ (MeV) &$T_{slope}$ (MeV) &$T_{slope}$ (MeV)  \\
&&(Ni+Al)&(Ni+Ni)&(Ni+Ag) \\
\hline

3&6&8.4&7.8&8.4 \\
3&7&7.6&9.0&8.2 \\
3&8&7.8&7.8&8.4 \\
4&7&9.5&9.7&10.8 \\
4&9&9.9&10.5&9.8 \\
4&10&10.7&9.7&12.1 \\
5&10&10.9&9.6&10.1 \\
5&11&10.4&10.0&10.9 \\
6&12&7.9&8.7&10.5 \\
6&13&7.4&9.1&10.5 \\
\hline
  \end{tabular}
\end{table}

\begin{table}
%\begin{center}
%{\bf Table II}
%\end{center}
\caption{
Isotopic composition of fragments emitted by the QP's (statistical errors 
are of the order of a few \%)}
  \begin{tabular}{lllll}
\hline
$Z$&$A$&Ni+Al (\%) &Ni+Ni (\%) &Ni+Ag (\%)  \\
\hline
3& 6&34.5&41.3&33.7 \\
3& 7&39.3&35.3&43.5 \\
3& 8&26.2&23.4&22.8 \\
4& 7&45.6&54.3&31.0 \\
4& 9&36.7&31.0&43.3 \\
4&10&17.7&14.7&25.7 \\
5&10&37.5&40.6&37.1 \\
5&11&52.2&48.1&50.1 \\
5&12&10.3&11.3&12.8 \\
6&11&13.6&17.8&15.6 \\
6&12&44.0&48.3&41.1 \\
6&13&34.8&25.2&31.1 \\
6&14& 7.6& 8.7&12.2 \\
\hline
  \end{tabular}
\end{table}

\begin{table}
%\begin{center}
%{\bf Table III}
%\end{center}
\caption{
Average N/Z values of the IMFs emitted from the QP}
  \begin{tabular}{llll}
\hline
$Z$&$<$N/Z$>$$_{qp(Ni+Al)}$&$<$N/Z$>$$_{qp(Ni+Ni)}$&$<$N/Z$>$$_{qp(Ni+Ag)}$ \\
3&1.31&1.27&1.30 \\
4&1.07&1.02&1.16 \\
5&1.15&1.14&1.15 \\
6&1.06&1.04&1.07 \\
\hline
  \end{tabular}
\end{table}

\begin{table}
%\begin{center}
%{\bf Table IV}
%\end{center}
\caption{Temperature and excitation energy of QPs} 
  \begin{tabular}{lll}
\hline
 &T (MeV)&E$^*$ (MeV/nucleon) \\
\hline

Ni+Al&3.7$\pm$0.2&4.0$\pm$0.5\\
Ni+Ni&3.9$\pm$0.2&4.0$\pm$0.5\\
Ni+Ag&4.1$\pm$0.2&3.5$\pm$0.5\\
\hline
  \end{tabular}
\end{table}

\begin{table}
%\begin{center}
%{\bf Table V}
%\end{center}
\caption{
Average N/Z values of the neck IMFs}
  \begin{tabular}{llll}
\hline
Z&$<$N/Z$>$$_{Neck(Ni+Ni)}$&$<$N/Z$>$$_{Neck(Ni+Ag)}$ \\
3&1.50&$\geq$1.44 \\
4&1.43&$\geq$1.36 \\
5&1.26&$\geq$1.26 \\
6&1.25&$\geq$1.18 \\
\hline
  \end{tabular}
\end{table}

%\end{multicols}
\end{document}